\documentclass[twocolumn, switch]{article} 

\usepackage{amsmath, amsthm, amssymb, amsfonts}

\usepackage[numbers,square]{natbib}
\usepackage[utf8]{inputenc}	
\usepackage{epstopdf} 
\usepackage{graphics}
\usepackage{placeins}
\usepackage{graphicx}
\usepackage[T1]{fontenc}	
\usepackage{xcolor}		
\usepackage[colorlinks = true,
            linkcolor = purple,
            urlcolor  = blue,
            citecolor = cyan,
            anchorcolor = black]{hyperref}	
\usepackage{booktabs} 		
\usepackage{nicefrac}		
\usepackage{microtype}		
\usepackage{lineno}		
\usepackage{float}
\restylefloat{table}			

\usepackage{lipsum}		

\usepackage{newfloat}
\DeclareFloatingEnvironment[name={Supplementary Figure}]{suppfigure}
\usepackage{sidecap}
\sidecaptionvpos{figure}{c}

\usepackage{titlesec}
\titlespacing\section{0pt}{12pt plus 3pt minus 3pt}{1pt plus 1pt minus 1pt}
\titlespacing\subsection{0pt}{10pt plus 3pt minus 3pt}{1pt plus 1pt minus 1pt}
\titlespacing\subsubsection{0pt}{8pt plus 3pt minus 3pt}{1pt plus 1pt minus 1pt}

\usepackage{tikz,xcolor,hyperref}

\definecolor{lime}{HTML}{A6CE39}
\DeclareRobustCommand{\orcidicon}{
	\begin{tikzpicture}
	\draw[lime, fill=lime] (0,0) 
	circle [radius=0.16] 
	node[white] {{\fontfamily{qag}\selectfont \tiny ID}};
	\draw[white, fill=white] (-0.0625,0.095) 
	circle [radius=0.007];
	\end{tikzpicture}
	\hspace{-2mm}
}
\foreach \x in {A, ..., Z}{\expandafter\xdef\csname orcid\x\endcsname{\noexpand\href{https://orcid.org/\csname orcidauthor\x\endcsname}
			{\noexpand\orcidicon}}
}

\title{Alarm-Based Root Cause Analysis in Industrial Processes Using Deep Learning}

\usepackage{xwatermark}
\newwatermark[firstpage,color=gray!60,angle=90,scale=0.32, xpos=-4.05in,ypos=0]{\href{https://doi.org/}{\color{gray}{Publication doi}}}
\newwatermark[firstpage,color=gray!60,angle=90,scale=0.32, xpos=3.9in,ypos=0]{\href{https://doi.org/}{\color{gray}{Preprint doi}}}
\newwatermark[firstpage,color=gray!90,angle=0,scale=0.28, xpos=0in,ypos=-5in]{*correspondence: \texttt{n.javanbakht712@yahoo.com}}

\usepackage{authblk}

\author{Negin Javanbakht}
\author{Amir Neshastegaran}
\author{Iman Izadi}

\affil{Department of Electrical and Computer Engineering, Isfahan University of Technology, Isfahan 84156-83111, Iran}

\begin{document}
\twocolumn[ 
  \begin{@twocolumnfalse} 
  
\maketitle

\begin{abstract}
 Alarm management systems have become indispensable in modern industry. Alarms inform the operator of abnormal situations, particularly in the case of equipment failures. Due to the interconnections between various parts of the system, each fault can affect other sections of the system operating normally. As a result, the fault propagates through faultless devices, increasing the number of alarms. Hence, the timely detection of the major fault that triggered the alarm by the operator can prevent the following consequences. However, due to the complexity of the system, it is often impossible to find precise relations between the underlying fault and the alarms. As a result, the operator needs support to make an appropriate decision immediately. Modeling alarms based on the historical alarm data can assist the operator in determining the root cause of the alarm. This research aims to model the relations between industrial
alarms using historical alarm data in the database. Firstly, alarm data is collected, and alarm tags are sequenced. Then, these sequences are converted to numerical vectors using word embedding. Next, a self-attention-based BiLSTM-CNN classifier is used to learn the structure and relevance between historical alarm data. After training the model, this model is used for online fault detection. Finally, as a case study, the proposed model is implemented in the well-known Tennessee Eastman process, and the results are presented.
\end{abstract}

\textbf{Keywords} Alarm management, \and Root cause analysis, \and Fault detection, \and Deep neural network, \and Word embedding, \and Tennessee Eastman process 
\vspace{0.35cm}

  \end{@twocolumnfalse} 
] 



\section{Introduction}
Industrial processes rely on SCADA systems for monitoring and controlling processes locally and remotely. The controllers communicate with field instrumentations such as sensors, pumps, and motors and collect real-time data. Then the data is processed, and various control algorithms are implemented to ensure those process variables remain within the acceptable range. If the implemented control does not work appropriately or an abnormal situation occurs in the process, the alarm system triggers an alarm to notify the operator of the abnormal conditions. 
According to ISA 18.2 standard ~\cite{isa2009isa}, an alarm is an audible or visual alert that informs the operator of abnormal situations such as equipment failures or process variables exceeding a specified threshold. When an alarm is raised, the operator must intervene and immediately find the root cause of the raised alarm. Otherwise, this fault can propagate through other equipment and affected variables, causing severe damage to the system and equipment and even stopping the whole process.

As processes become larger and more interconnected, faults in one component can affect other equipment and variables. Hence, faults can propagate through faultless parts of the plants and lead them to abnormal conditions, and as a result, alarm numbers increase considerably~\cite{ahmed2013similarity}. Accordingly, the operator is confronted with an enormous number of alarms which make him overwhelmed; hence, the operator is unable to handle the abnormal situation properly. Therefore, it is of high importance to find the root cause of the fault as soon as possible to confine or prevent further events.

 Because of equipment failures, faults are inevitable in industrial processes. When an alarm is raised, it is likely to be a sign of fault occurrence. As a result, alarms are our guidelines for fault detection. Due to the system's complexity and interactions, there is no one-to-one relation between alarms and faults. Thus, a mechanism is required that assists the operator to the occurred fault according to alarms.
Various researchers have so far addressed root cause analysis using process variables. Methods such as transfer entropy ~\cite{hu2017cause} and modeling based on deep learning methods such as convolutional neural networks ~\cite{wu2018deep} have been used for this issue. However, one can take advantage of alarm data instead of employing process data. Process data results from constantly measuring process variables; thus, the data volume is large, and data access is difficult. This issue increases the amount of the calculations. 
On the other hand, because process data is frequently generated, the algorithms must be run continuously for online implementation. In contrast, alarm data is generated only in case of a fault occurrence or exceeding a process variable from pre-defined thresholds. Therefore, an online algorithm for alarm data needs to be run only in the event of such problems.

Root cause analysis using alarm data has not extensively been addressed. In a related study, patterns are extracted based on the relations between alarms, used for online fault detection. This approach is implemented in ~\cite{charbonnier2016fault} as follows. At first, sequences and vector patterns of faults are extracted; then, these patterns are combined to form a sequential weighted fault pattern and are used as a detection criterion. Pattern extraction in ~\cite{noroozifar2019root} is done by creating an initial presence/absence matrix based on the relationships between alarms and faults. 

 Another approach is to model the relations between the alarms. One method employs a causal model to represent causal dependencies between alarms and determine the alarms' root cause. The causal dependencies among alarms are modeled via a Bayesian network, as a causal model in ~\cite{wunderlich2017structure}. There are several approaches to learning the structure of a Bayesian network. To determine the best method, in ~\cite{wunderlich2017structure}, one algorithm from each method is used to find the best method for modeling alarm data, and the results are compared. The extracted model is compared to the exact model of the system to select the best model of the dependencies between alarms. As a result, basic system information is required to obtain the correct model.

Another causal model used for modeling alarm relations is the Petri net, which is used in ~\cite{no}. A Petri net model is used for each fault scenario, to represent the relations between alarms and their patterns. These models are evaluated based on criteria specified in the process mining literature to determine their ability to represent each fault scenario. When a fault occurs, corresponding alarm tags are compared to the extracted Petri net models to detect the underlying fault scenario.
The process mining method used in this paper is only applicable to the fault scenarios in which the relational patterns between alarm tags can be described using sequence operators, exclusive selection, parallel execution, and loop execution. In contrast, neural networks can extract a broader range of patterns using nonlinear functions. Therefore, more fault scenarios can be detected using neural networks.
 
Deep learning structures and neural networks are other methods for modeling alarm dependencies. Applying the neural network with LSTM layers for root cause analysis of alarm data is the approach employed in ~\cite{dorgo2018understanding}. The approach presented in this paper is intended to improve the performance of this deep learning model by modifying the structure of the neural network. The improved neural network must fulfill three criteria. The neural network must be trained using a small amount of data while maintaining perfect precision and accuracy. Besides, the learning procedure must be incredibly fast. These three criteria are critical in online fault detection. In industries, there aren't a lot of fault scenarios, so we tried to keep the volume of data required for training a model as small as possible in this research. Besides, in online fault detection, the operator must immediately determine the root cause of the alarm. So, the detection time is a significant factor. Therefore, the current paper aims to present a fast neural network training procedure that assists the operator in quickly determining the root cause of an alarm. The contents of this paper are summarized below:

\begin{enumerate}
\item
Alarm data needs to be preprocessed. First, chattering alarms are removed, and then alarm tags are organized in sequences.
\item
Since the neural network input must be numerical vectors, alarm data in the textual format is first converted into numerical vectors. Then a window of length $ v $ moves on these vectors, and each sequence is segmented into some subsequences.
\item
Generated subsequences are randomly divided into train, validation, and test sets, and the neural network is trained with this data and then is tested.
\item
After training, the neural network is used for online fault detection.
\end{enumerate} 

The remainder of this paper is structured as follows. The layers used in the proposed network structure are introduced in section 2. Section 3 discusses alarm management systems and their alarm data, as well as a brief introduction to alarm data preprocessing. Section 4 is devoted to the designed neural network structure, and Section 5 focuses on implementing this structure as a case study on the Tennessee Eastman process. Section 6 offers a conclusion.

\section{Deep learning layers for the proposed structure}\label{}
In this section, the one-dimensional Convolution layer, the Bidirectional LSTM layer, and the Self-Attention layer, used for the structure of the proposed model for fault detection, are introduced.

\subsection{One-dimensional convolution layer}
A single one-dimensional convolution layer is used in the proposed model to extract features from input sequences and reduce their size.  
By choosing the most important features, the size of the input matrix decreases significantly, so the computational cost decreases. As a result, this layer can speed up the neural network training procedure. This layer consists of $n$ one-dimensional filters with variable window sizes (kernel size).

Input sequences of the neural network must be numerical vectors. To use textual sequences as an input of a CNN layer, first, these sequences must be converted into numerical vectors using Word embedding ~\cite{cai2019process}. Then CNN filters move on these input sequences to extract features. The height of the filter is window size and is selectable (kernel size), and its length equals the embedding dimension of input sequences.
 Input sequences are multiplied by the filter's weights element-wise, and then this passes to a nonlinear activation function (relu)  and the features of the input sequences are extracted as sequences ~\cite{liu2019bidirectional}. Eq. \ref{l} shows this concept \cite{liu2019bidirectional}:
\begin{equation}
\label{l}
Lc_{n}=f(w\cdot x_{i:i+m-1}+b)
\end{equation}
where $w$ is the weight matrix of the CNN layer, $b$ denotes the bias vector. Also $ x_{i:i+m-1} $
 is the word embedding representation of $m$ words starting from the $i-th$ word and $Lc_{n}$ is the $n-th$ feature sequence generated ~\cite{liu2019bidirectional}.
 
Considering the number of filters to be j, the feature sequences extracted from words $ x_{i:i+m-1} $ are:
$Lc=[Lc_{1}, Lc_{2},\ldots ,Lc_{j}]$.

\begin{figure*}[t]
 \centering
\includegraphics[scale=0.7]{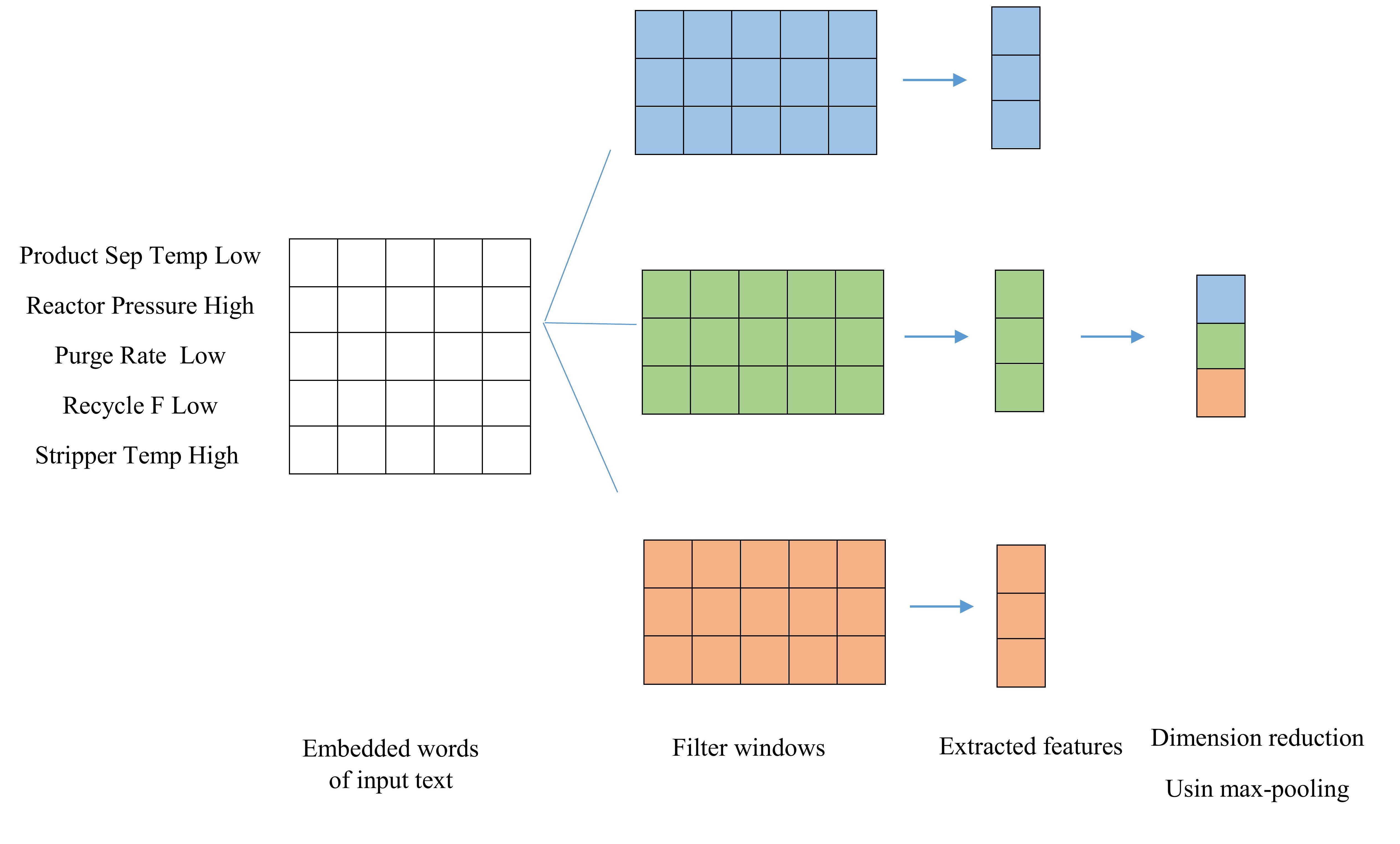}
\caption{Architecture of Convolutional layer \cite{liu2019bidirectional}}
\label{c}
\centering
\end{figure*}

Having extracted feature sequences, it is possible to reduce the size of matrices using a max-pooling layer. Fig.~\ref{c} shows an overview of the convolutional layer.

\subsection{BiLSTM layer}
\begin{figure*}[t]
 \centering
\includegraphics[scale=1]{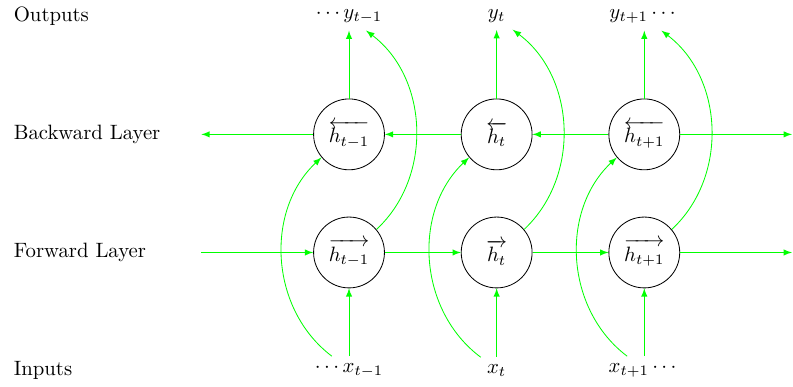}
\caption{Architecture of BiLSTM layer \cite{li2020bidirectional}}
\label{a}
\centering
\end{figure*}

LSTM units are capable of learning dependencies between the tokens of sequences. As it is desirable to capture the dependencies in the alarm sequences, LSTMs are proper alternatives to be selected. In the structure of each LSTM cell, there are hidden state $h_{t}$ and memory cell $c_{t}$. In respect with this unit, hidden states $h_{t}$ and memory cell $c_{t}$ are computed based on the previous states $h_{t-1}$, $c_{t-1}$, and input vector $ x_{t} $. LSTMs, considering the input context only  in the forward direction, is given as \cite{li2020bidirectional}:
\begin{equation}
c_{t}, h_{t}=g^{lstm}(c_{t-1},h_{t-1}, x_{t})
\end{equation}
Bidirectional LSTM contains the forward LSTM and the backward LSTM simultaneously and considers the context in both forward and backward directions. As a result, they considerably outperform LSTMs in extracting the dependencies. Forward LSTM $(\overrightarrow {LSTM})$ captures semantic dependencies in the forward direction, and the output is forward hidden state sequence $(\overrightarrow {h(t)})$. Backward LSTM $ (\overleftarrow{LSTM}) $ does the same function in backward direction and backward hidden state sequence $ (\overleftarrow{h(t)}) $ is the output. Then hidden states of two LSTMs are concatenated, and a Bidirectional LSTM hidden state is formed. The function of the BiLSTM layer is illustrated in Fig.~\ref{a} and formulated as follows \cite{li2020bidirectional}:
\begin{equation}
\overrightarrow {c_{t}}, \overrightarrow {h_{t}}=g^{lstm}(\overrightarrow {c_{t-1}},\overrightarrow {h_{t-1}}, x_{t})
\end{equation}
\begin{equation}
\overleftarrow {c_{t}},\overleftarrow {h_{t}}=g^{lstm}(\overleftarrow {c_{t+1}},\overleftarrow {h_{t+1}
}, x_{t})
\end{equation}
\begin{equation}
h_{t}=\overrightarrow {h_{t}}\oplus \overleftarrow {h_{t}}
\end{equation}

\subsection{Self-attention layer}
The purpose of using the Attention layer is to focus on important concepts and attend to decisive information. By using the Attention mechanism, we can emphasize the significant tokens in an input sequence ~\cite{zheng2018opentag}. The Self-Attention layer gives different weights to the vectors of individual words in a sentence, depending on their similarity of adjacent tokens ~\cite{polignano2019comparison}. This layer is added following the BiLSTM layer. Thus, hidden state sequences $h_{t}$ are inputs of the Self-Attention layer. The similarity between hidden states $ h_{t} $ and $h_{t\ensuremath{'}}$ at timestamps $ t $ and $ t\ensuremath{'} $ is captured by $ a_{t,t\ensuremath{'}} $ ~\cite{zheng2018opentag}. The importance of a token at any timestamp according to the neighboring context is represented by $l_{t}$. It is defined by the weighted sum of hidden state $h_{t\ensuremath{'}}$ of all other tokens at timestamp ${t\ensuremath{'}}$. This can be useful in determining the final class of the model ~\cite{zheng2018opentag}.
These mathematical relations are formulated as below ~\cite{polignano2019comparison}:
\begin{equation}
g_{t,t\ensuremath{'}}=\tanh (W_{g}h_{t}+W\ensuremath{'}_{g}h_{t\ensuremath{'}}+b_{g})
\end{equation}
\begin{equation}
\alpha _{t,t\ensuremath{'}}=\sigma (W_{\alpha}g_{t,t\ensuremath{'}}+b_{\alpha})
\end{equation}
\begin{equation}
a _{t,t\ensuremath{'}}=softmax(\alpha _{t,t\ensuremath{'}})
\end{equation}
\begin{equation}
l _{t}=\sum ^{n}_{t\ensuremath{'}=1} a _{t,t\ensuremath{'}}.h_{t\ensuremath{'}}
\end{equation}
where $W_{g}$ and $W\ensuremath{'}_{g}$ are weight matrices relating to hidden states $h_{t}$ and $h_{t\ensuremath{'}}$, $W_{\alpha}$ shows their non-linear combination weight matrix, and $b_{g}$ and $b_{\alpha}$ are bias vectors ~\cite{zheng2018opentag}.

Using this layer after the BiLSTM layer instead of another BiLSTM reduces the network's computations and speeds up the training procedure. This layer's output is weighted over its input sequences. The weighting of the input sequences is determined by the tokens that are adjacent to them in each sequence. These calculations improve the model's accuracy and precision in predicting the final class of each input sequence.

\subsection{Overfitting}
\begin{figure}[t]
 \centering
\includegraphics[scale=0.7]{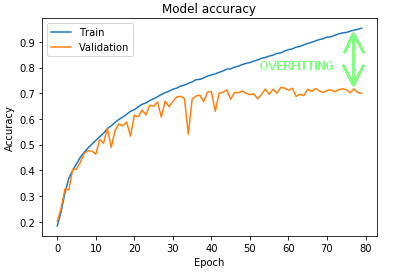}
\caption{Overfitting on a model}
\label{aqq}
\centering
\end{figure}

\begin{figure}[t]
 \centering
\includegraphics[scale=0.6]{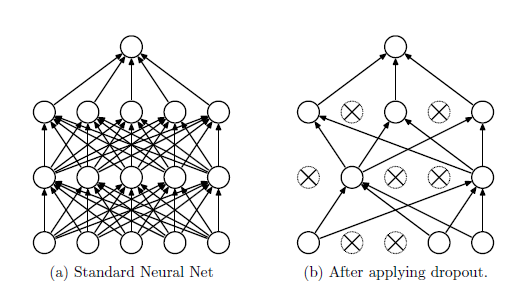}
\caption{Dropout concept \cite{dropout}}
\label{wwe}
\centering
\end{figure}

The purpose of training a neural network is to find a model that fits well on both training and validation data. The model extracted by neural network using training data should be generalized to the new data. Overfitting occurs when the model is well trained on the training data but does not perform well on the unseen data. Figure \ref{aqq} depicts overfitting on a model. There are numerous methods to prevent overfitting. Dropout is the method used in this paper to avoid overfitting. In this method, some of the neurons of the neural network are removed randomly. Every node can be removed with the p probability and retained with the 1-p probability. As a result, learning occurs on various architectures with various sets of neurons, and the final output is formed by combining the outputs of several networks \cite{dropout}. The dropout concept is depicted in Fig. \ref{wwe}. 

\section{Alarm management and data preprocessing}
The following sections discuss the fundamental principles of the alarm management system, alarm data, and alarm data preprocessing.

\subsection{Alarm system}
\begin{figure}[t]
 \centering
\includegraphics[scale=0.5]{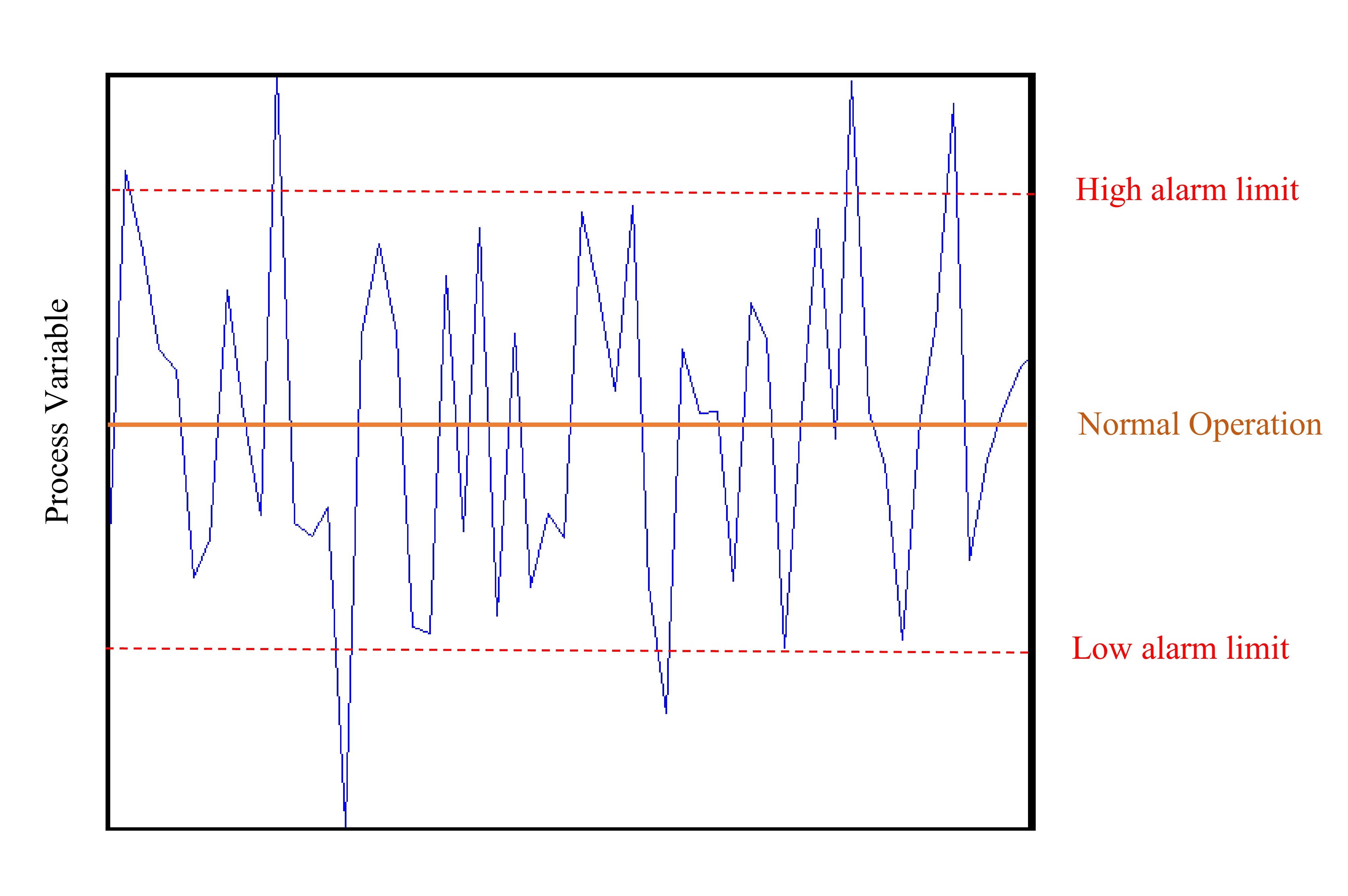}
\caption{Basic alarm activation method ~\cite{izadi2010effective}}
\label{cnn2}
\centering
\end{figure}

An acceptable range is defined for each variable's behavior in the plants and processes $(e.g., \enspace \mu \pm{3\sigma})$. If the controlled variable exceeds these thresholds, an alarm is triggered ~\cite{izadi2010effective}. This is the simplest way of triggering an alarm shown in Fig. \ref{cnn2}. Also, alarm systems were designed to raise an alarm when a fault occurs. In this case, the fault can propagate through other equipment and affected variables, leading to an enormous number of warnings and alarms. This highlights the importance of finding the root cause of an alarm immediately.

\subsection{Alarm data}
Generated alarms are text messages or are saved in the database. Each line of alarm data contains information, including:

Timestamp: time of triggering an alarm.

Tag: process variable that raised an alarm.

Alarm identifier: identifies the type of alarm (e.g., low alarm, low low alarm, high alarm, high high alarm, etc.).
 
\begin{figure*}
\includegraphics[scale=0.8]{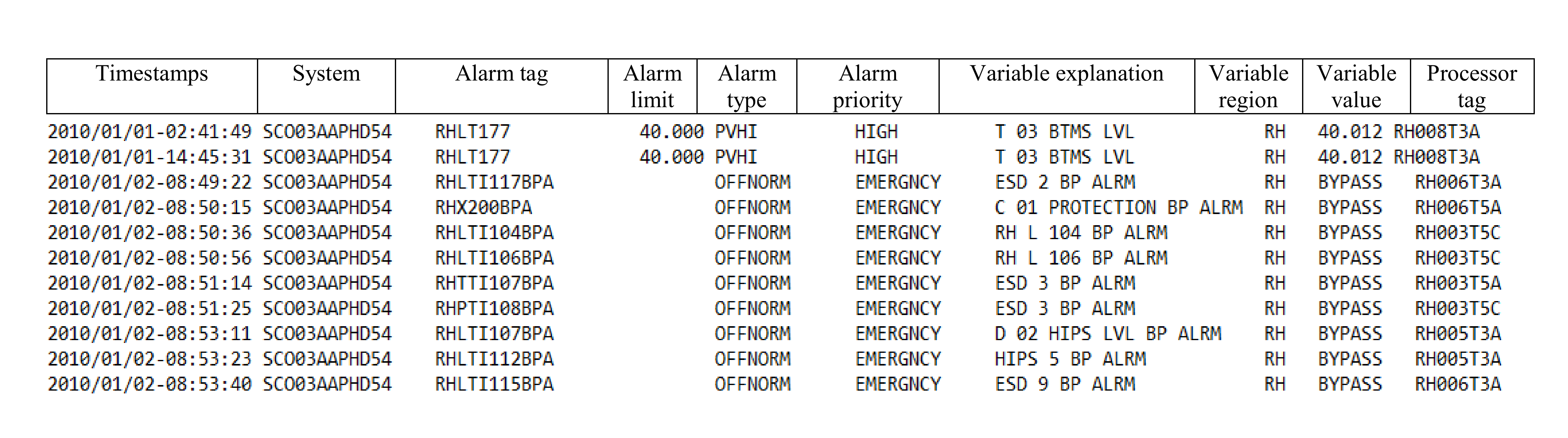}
\caption{A sample of raw alarm data}
\label{n}
\end{figure*}

Priority: shows the emergency  and seriousness of an alarm.
Fig. \ref{n} shows a sample of raw alarm data.

\subsection{Word embedding}
Alarm tags (e.g., Reactor Coolant Temp) and tag identifiers (e.g., High) are used to extract alarm sequences. Since the neural network inputs must be numerical vectors, these extracted sequences must be converted into numerical vectors. One approach is to use one-hot encoding. Each unique alarm tag is represented as a unique binary-element vector, and repeated tags are encoded using this unique representation. Thus if the total number of alarm tags is $n$ and the number of unique alarm tags is $ p$, then the created matrix's size is $ n\times p $. This matrix is large-sized and sparse. 

Another option is to employ Word2Vec ~\cite{3}. In this case, each unique alarm tag has its numerical representation vector of length $r$. Hence, the embedding matrix is $ n\times r $ where $n$ is the number of total alarm tags. The length of the vectors decreases significantly  using Word2Vec, and as a result, the volume of the calculations is reduced. Word2Vec was chosen as the method for word embedding in this paper. 

Word2Vec is a two-layer feedforward neural network that takes a text in the form of sequences, whose output after training is a matrix. Each row of this matrix is the representation of a unique word in the input text. The architecture used for Word2Vec  in this paper is Skip-Gram.

\subsection{Alarm data preprocessing}
The data used for training a neural network must be preprocessed and converted to an appropriate format. Our objective of training a neural network is to identify the fault scenario according to alarm data. In other words, this neural network identifies the fault scenario as an output for an input sequence of alarm data. Hence, alarm data must be labeled according to the respective fault. This task could be done by maintenance and repair reports and daily event records. However, it is not possible to label alarm data accurately. Therefore, the amount of data available for training a neural network is limited. 

It is possible to reduce input complexity to compensate for the amount of alarm data used in training a neural network. In other words, alarm data can be selected in a way that alarm sequences associated with a fault scenario are more similar to one another while remaining distinct from alarm sequences associated with other fault scenarios. Primary alarms triggered after fault occurrence are very likely to accord with this feature. Each fault starts at one point or area and, if left unchecked, can propagate over time. Different faults typically originate in various parts of the process, resulting in distinct alarms. However, over time and as the fault propagates, the possibility of finding similar patterns in their behavior increases. Besides, the faster the type of fault is detected from the first alarms, the more ideal it is because it can be fixed faster or prevented from propagating in the process. As a result of the aforementioned reasons, the first $ k $ alarms were used for fault detection rather than the entire alarms of each fault scenario.

In this section, alarm tags of each fault are placed in a sequence chronologically. The first $k$ alarm tags of each sequence are then retained, while the rest are removed. The larger is the $  k$, the network is trained for the more comprehensive data. In contrast by raising k, data complexity increases, and network accuracy may decrease due to the limited amount of training data. Therefore, for online fault detection, a compromise must be made between the comprehensiveness of the network and its accuracy. In the following, these sequences are labeled with the corresponding fault scenario. As a result, the alarm database is converted to a set of labeled alarm sequences with $k$ length.

 The trained neural network must identify the fault scenario by observing $v$ consecutive alarm tags from the $k$ tags in each sequence. In fact, $v$ is the number of alarms needed to detect the fault online. In choosing $v$, the lower the value of $v$, the lower the delay is, which is naturally closer to ideal. However, by doing so, the probability of forming identical alarm sequences with different labels increases. As a result, the network estimation performance decreases because the input data (alarm sequences) of other classes (fault scenarios) are not sufficiently differentiated. Increasing the length of $v$ can reduce the probability of forming such sequences, which improves the network's performance. However, this will come at the expense of increasing the number of required alarms for neural network input, resulting in a delay in detecting the fault. Therefore, a compromise must be made between improving network performance and delay in fault detection.
 
First, all alarm tags are converted into numerical vectors using the Word2Vec model. Then a window with a length of $v$ is swept on each $k$ length sequence to form $v$ length sequences out of them.
Hence, from each primary $k$ length sequence, $k-v+1$ secondary sequences of size $v$ are generated. The labels of secondary sequences are similar to initial sequences. These secondary sequences are neural network input, and their labels (fault scenario) are neural network output. In fact, this neural network is supposed to classify $v$ length input sequences according to the fault scenario.

\section{Proposed neural network architecture}
The overview of the designed classifier for fault detection is discussed in this section.

\subsection{Alarm data  classification}
Alarm sequences generated by a common fault are classified into a class by designing a classifier. In the ideal case, each alarm sequence belongs to only one class; thus, the data type is considered categorical, and the number of classes equals the number of fault scenarios. Hence fault classes, $y_{i}= \{1, 2, \cdots , i\} $ where $ i $ indicates the number of classes, must be one-hot encoded.

\subsection{Neural network architecture}
The neural network proposed for this classification problem is a six-layer neural network, including input layer, one dimensional CNN, BiLSTM, Self-Attention, Dense layer and output layer.

\begin{figure*}[t]
\centering
\includegraphics[scale=0.6]{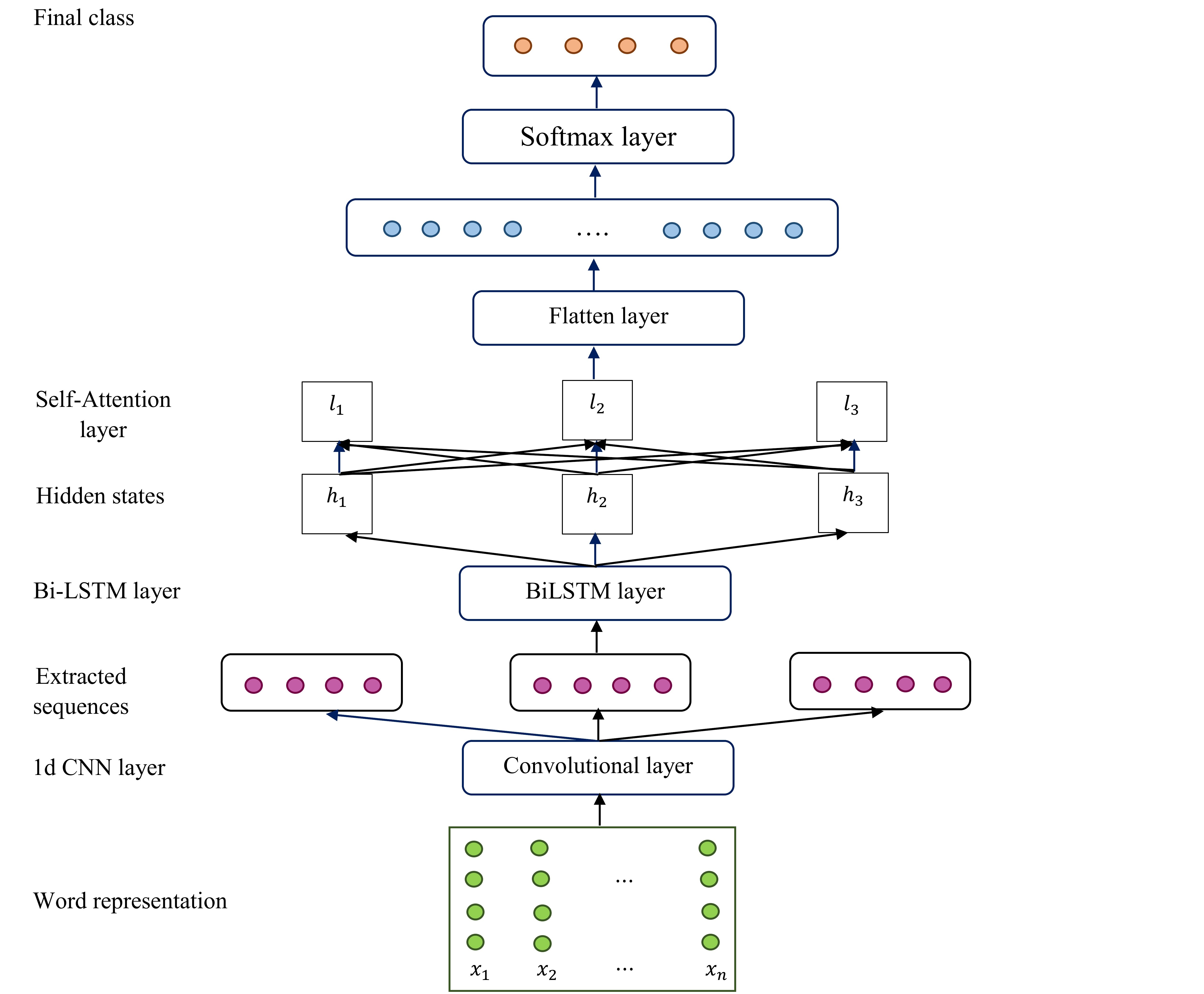}
\caption{Model structure for fault detection \cite{liu2019bidirectional}}
\label{u}
\centering
\end{figure*}

\begin{enumerate}
\item
The first layer is the input layer, where alarm sequences are given to the neural network.
\item
A one-dimensional CNN layer serves as the second layer. This layer is used to extract features from input sequences. As explained in section 2, input features are extracted as sequences in this layer. Furthermore, it significantly reduces the input dimension. 
\item
The sequences extracted from the second layer are the input of the BiLSTM layer. The outputs of this layer are hidden state sequences $ h_{t} $.
\item
The fourth layer is a Self-Attention layer. The hidden states of the previous layer are the input of this layer.
After the Self-Attention layer, a Flatten layer converts a 3-dimensional matrix to a 2-dimensional one.
\item
The fifth layer is the Dense or Fully-Connected layer. This layer has a softmax activation function and determines the probability of its input sequences being the network's final class. 
\item
The fault scenario corresponds to the input sequences determines the model's output in the final layer. 
\end{enumerate}
The architecture of this deep learning model is shown in Fig. \ref{u}.

\subsection{Training neural network using alarm data}
\begin{figure*}[t]
\centering
\includegraphics[scale=1]{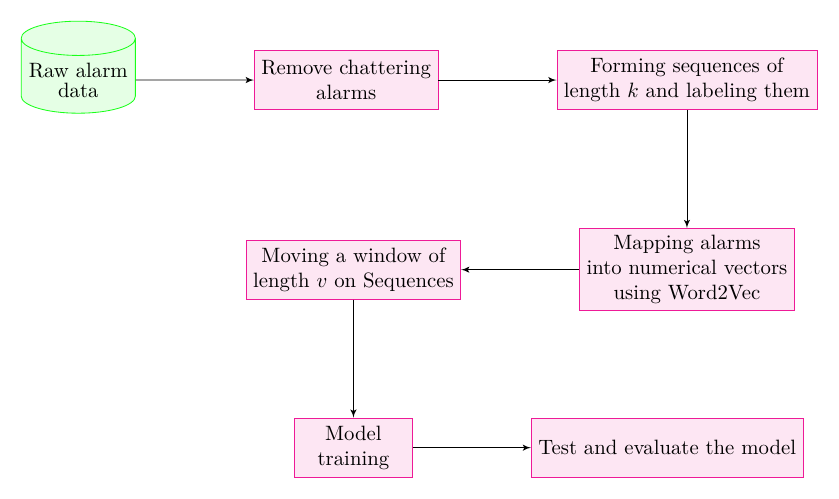}
\caption{Procedure of training a neural network}
\label{mo}
\centering
\end{figure*}

Alarm tags and tag identifiers, such as Reactor Coolant Temp+ High, are used to train a neural network. As previously discussed, text data must be converted to numerical vectors before being used to train a neural network. The first $ k=20 $ alarms in each sequence are retained; then, these sequences are given to the Wored2Vec model. Word2Vec is implemented with the gensim library. The output of the Word2Vec model is numerical vectors. Following the formation of numerical vectors, a window of length $v=5$ sweeps on these created vectors, and the data is segmented into five-length subsequences. These subsequences are randomly divided into train, validation, and test data. The output vector is the one-hot encoded fault scenario associated with each five-length subsequence. These steps are illustrated in Fig. \ref{mo}.

\subsection{Online fault detection}
The purpose of modeling alarm data using neural networks is to use it for fault detection when a fault occurs. As a result, in this section, an algorithm for online fault detection using the proposed model is presented.

\begin{figure*}
\includegraphics[scale=1.2]{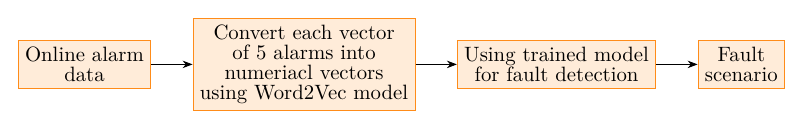}
\caption{Suggested approach for online fault detection}
\label{rt}
\end{figure*}

\begin{enumerate}
\item
 When five consecutive alarms are raised, this alarm sequence is converted to numerical vectors using the previously obtained Word2Vec vectors. 
\item
These numerical vectors are then transformed into a matrix, and the matrix is assigned to the trained model.
\item
The output of the model is the fault scenario.
\end{enumerate}
This procedure is summarized in Fig.~\ref{rt}.

\section{Case study}
In this section, the proposed neural network is implemented to analyze alarm data of the Tennessee Eastman process.

\subsection{Alarm management for the Tennessee-Eastman process}
\begin{figure*}
\includegraphics[scale=0.6]{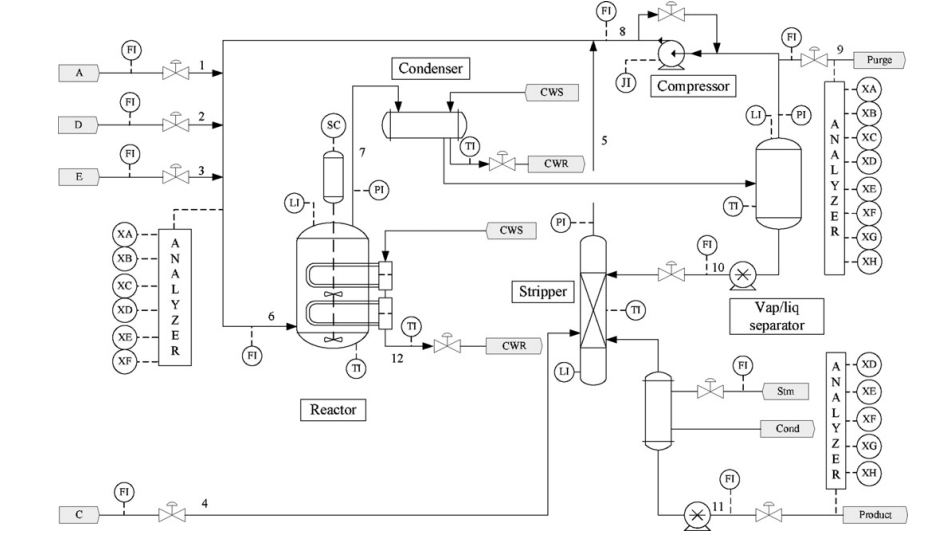}
\caption{Diagram of the Tennessee Eastman Process ~\cite{1}}
\label{tennn}
\end{figure*}

Tennessee Eastman process ~\cite{1} is a chemical process simulation benchmark. Tennessee Eastman's company has simulated this process based on the actual process behavior. This simulation is used as a well-known criterion for evaluating process control and monitoring methods. 
This process consists of five operating units:  reactor, condenser, recycle compressor, separator, and stripper. Four reactants A, C, D, and E react with each other, and G and H are produced. Figure \ref{tennn} depicts a diagram of the Tennessee Eastman process.

The goal of this process is to achieve and maintain a specific mass ratio between products G and H at the output. The process has several operating modes depending on the output substance ratio. The mass ratios of G and H are equivalent in this paper's working mode. In this process, 41 process variables are measured, and 20 fault scenarios are planned.

However, this process lacks an alarm management system. An alarm management system was designed to record alarm data of this process at the time of fault occurrence. For each process variable with an average $ \mu$ and standard deviation $\sigma$ in the normal state, high alarm limits $ \mu +3\sigma $ and low alarm limits $ \mu-3\sigma $ are generally considered. However,  the alarm limits may vary, in some cases, to reduce false and missed alarm rates. The total number of alarm tags will be 82 considering low and high alarm limits for each of the 41 process variables.

\begin{figure*}[t]
\centering
\includegraphics[scale=0.7]{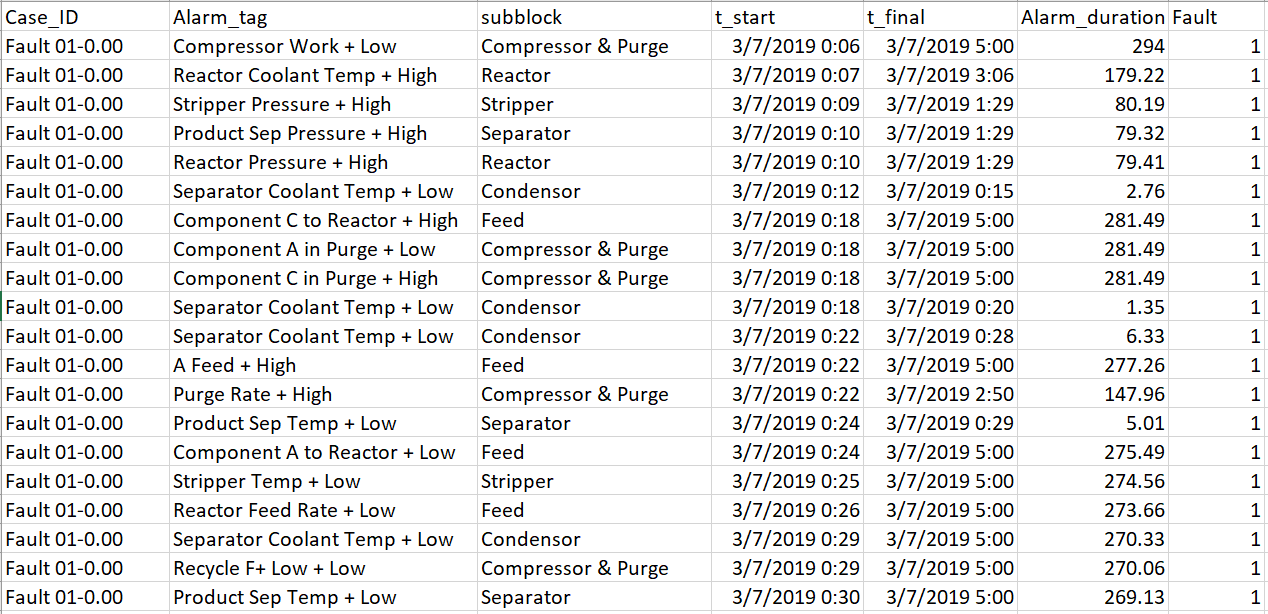}
\caption{Tennessee Eastman alarm data}
\label{b}
\end{figure*}

Chattering alarms are a common problem in alarm management systems. This phenomenon can be caused by reasons such as the nature of the process, noise, etc. To solve this issue, in designing the alarm management system, delay timers are applied according to ISA 18.2 in the alarm management system  ~\cite{2}. According to this standard, the delay for pressure and flow variables is 15 seconds and for temperature and level variables is 60 seconds.
Fig. \ref{b} shows a sample of Tennessee Eastman alarm data.

\subsection{Formation of input sequences}
For training the neural network, a dataset including records of 2000 alarms of the Tennessee Eastman process is used. The alarm data is related to fault scenarios 1, 2, 6, 7, 8, 10, 11, 12, 13, 17 obtained from 10 execution of each fault (100 executions in total).
Table \ref{hj} provides a brief description of the fault scenarios ~\cite{1}.
\begin{table*}[t]
\centering 
\caption{Process faults for the Tennessee Eastman process ~\cite{1}}  
\label{hj}
\begin{tabular}{@{}ccc}
\toprule
No.
&
Process variable
&
Type\\
\midrule 
1
& 
A/C feed ratio, B composition constant (stream 4)
& 
Step
\\
2
&
B composition, A/C feed ratio constant (stream 4) 
& 
Step

\\
3
&
D feed temperature (stream 2) 
& 
Step 
\\
4
&  
Reactor cooling water inlet temperature
& 
Step 
\\
5
& 
Condenser cooling water inlet temperature
& 
Step 
\\
6
& 
A feed loss (stream 1)
& 
Step
\\
7
& 
C header pressure loss-reduced availability (stream 4)
& 
Step
\\
8
&
A, B, and C feed composition (stream 4)
& 
Random variation
\\ 
9
&
D feed temperature (stream 2)
& 
Random variation 
\\ 
10
&
C feed temperature (stream 4)
& 
Random variation
\\ 
11
&
Reactor cooling water inlet temperature
& 
Random variation
\\ 
12
&
Condenser cooling water inlet temperature
& 
Random variation
\\ 
13
&
Reaction kinetics
& 
Slow drift
\\
14
&
Reactor cooling water value
& 
Sticking
\\
15
&
Condenser cooling water value
& 
Sticking
\\
16
&
Unknown
& 
Unknown
\\
17
&
Unknown
& 
Unknown
\\
18
&
Unknown
& 
Unknown
\\
19
&
Unknown
& 
Unknown\\
20
&
Unknown
& 
Unknown
\\
\bottomrule
\end{tabular}
\end{table*}
To train the network, alarm data related to the execution of each fault, including 20 alarms, are given to the Word2Vec model, and numerical vectors corresponding to each alarm tag are extracted. Then, a window of length 5 moves on each sequence. As a result, 16 subsequences with length 5 are created from each sequence. Therefore, 1600 subsequences of 5 alarm tags are created as network inputs. Moreover, corresponding fault scenarios of each alarm subsequences turn into one-hot encoded vectors and are the network output. The data is then randomly divided into the train, validation, and test sets. Test data is used to evaluate network performance after training.

\subsection{Implementation}
The proposed network overview presented in section 4.2 is implemented in Python environment using Keras library. Table \ref{h} illustrates the parameters of the implemented network. These parameters are tuned through trial and error. The parameters are modified multiple times, and the effect of their values is observed each time. The best values are chosen.

\begin{table*}[h!]
\centering
\caption{Parameters of the implemented network} 
\label{h}
\begin{tabular}{@{}cc@{}}
\toprule
Parameter
&
Value
\\
\midrule
Number of filters in the convolution layer
& 128
\\
Height of the 1D convolution window (kernel size)
&
3
\\
Number of BiLSTM units
& 
128
\\
Learning rate
&
0.00001
\\
Number of training epoch
& 
100 
\\
Batch size
& 
100
\\
Word embedding size
& 
10
\\
Dropout
& 
0.1
\\
\bottomrule
\end{tabular}
\end{table*}

\subsection{Results}
\begin{figure*}[t]
\centering
\includegraphics[scale=0.7]{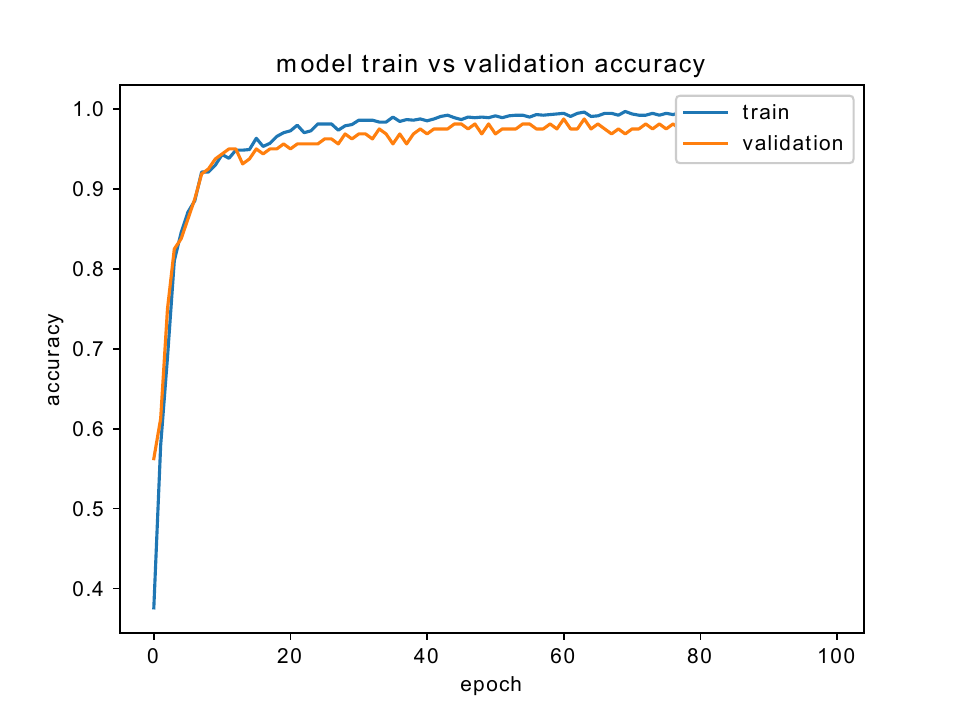}
\caption{Train and Validation accuracy}
\label{ty}
\centering
\end{figure*}

\begin{figure*}[h!]
\centering
\includegraphics[scale=0.6]{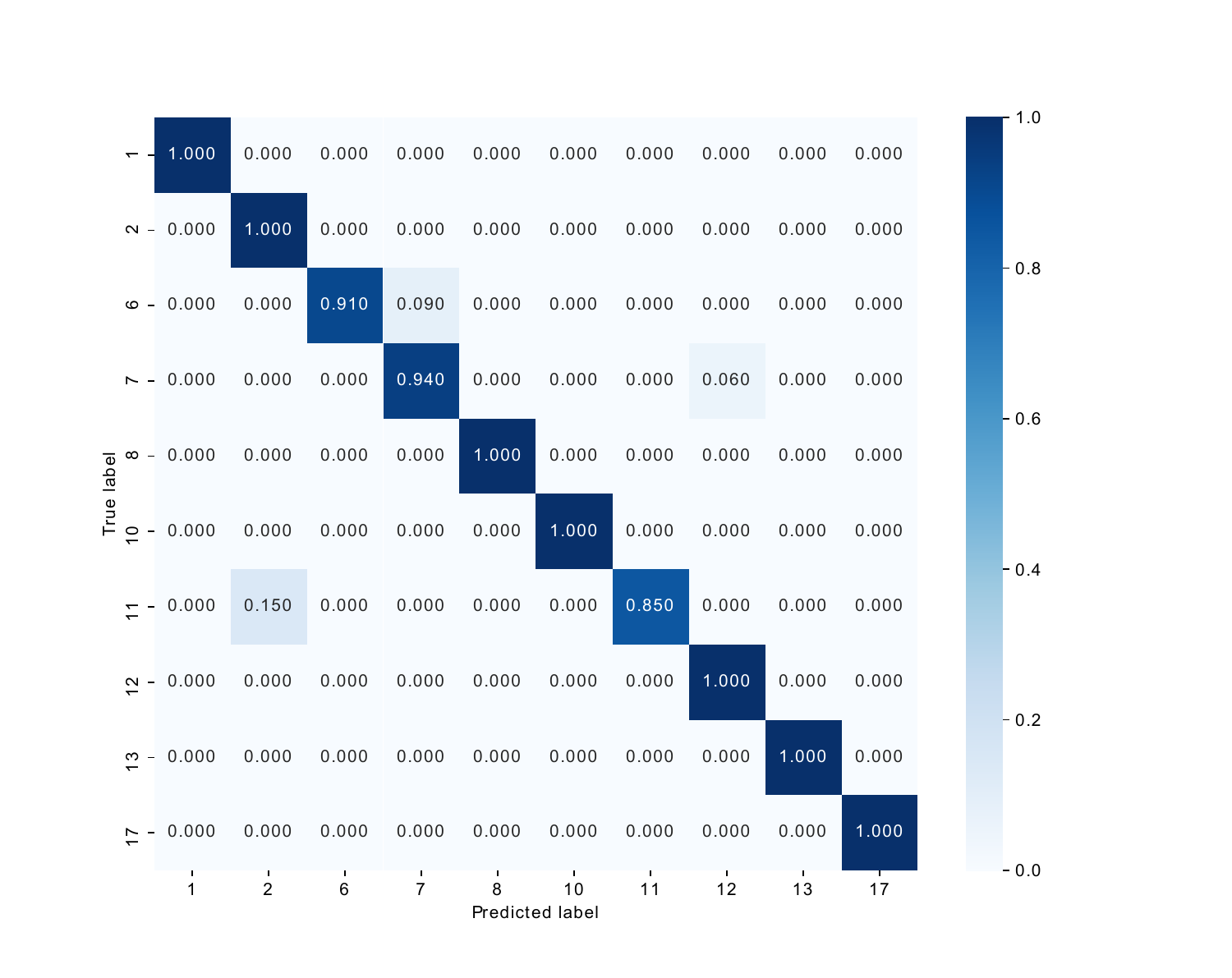}
\caption{Precision}
\label{ui}
\centering
\end{figure*}

The results of the implemented method are reported in this section. Fig \ref{ty} shows train and validation accuracy during the training procedure. The precision determined using the confusion matrix is presented in Fig. \ref{ui}. All the three criteria proposed in the introduction part of this paper are satisfied. The model is trained using 2000 alarm tags, which is a small amount of data compared to existing methods. Also, the accuracy and precision are perfect. Training this model takes about 22 seconds. Accordingly, this is an ideal model with a fast training procedure and excellent accuracy and precision, using only 2000 alarm tags.

\begin{table*}[t]
\centering
\caption{Results of the implemented network} 
\label{h2}
\begin{tabular}{@{}ccc@{}}
\toprule
Train accuracy
&
Validation accuracy
&
Test accuracy
\\
\midrule
0.99
&
0.98
&
0.96
\\
 \bottomrule
\end{tabular}
\end{table*}

\subsection{Online fault detection}
After training the neural network, the extracted model should be examined whether it can identify the fault scenarios by the first five alarm occurrences or not.
For this purpose, the first occurrence of each fault scenario containing $ k=20 $ alarms is selected. This selection is because the model should identify fault scenarios with the first fault occurrences. In this case, the operator can find the alarm's root cause as soon as possible. These sequences are converted to numerical vectors using Word2Vec vectors previously obtained from the trained Word2Vec model.
Then, these numerical sequences are swept by a window length $ v=5 $, and subsequences of length five are given to the trained model one by one. All the faults were detected correctly by the first five alarms. As time is a significant factor in fault detection, finding the root cause of the alarms by the first five alarms of each fault scenario is perfect and shows the model efficiency. 

\subsection{Comparison with previous works}
To demonstrate the superiority of the method of this paper over the existing approach that uses the Tennessee Eastman's alarm data for root cause analysis of alarm data ~\cite{no} using process mining methods, we will discuss the results of both methods.
In ~\cite{no} process mining method is used for extracting the model of the appeared alarms. These models are used for online fault detection after selecting the appropriate model according to the defined criteria. However, the process mining approach can only be used for the fault scenarios where the patterns between alarm tags can be described using sequence operators, exclusive selection, parallel execution, and loop execution. Hence, using the process mining approach, the models of fault scenarios 1, 2, 6, 7, and 11 are extracted. Alternatively, neural networks can extract a broader range of patterns using nonlinear functions. Therefore, more fault scenarios can be detected using neural networks. 

Another advantage of this method over the method in ~\cite{no} is that each sequence of five consecutive alarms may be utilized as neural network input in the proposed method. Whereas only the first five alarms of each fault scenario can be used for fault detection in ~\cite{no}.

A dataset including 15 occurrences of fault scenarios 1 and 2 is investigated to compare the results of the online fault detection of the model provided in this study with the model in ~\cite{no}. The first five alarms of each fault scenario are given to the model proposed in the current paper and the model presented in ~\cite{no}. The results related to ~\cite{no} are provided in tables \ref{nm} and \ref{op}. The results of the approach employed in this research are shown in tables \ref{cas} and \ref{cat}. Each row corresponds to a sequence of five alarms, and the associated fault scenario is determined in the column. The numbers in these tables represent the predicted probabilities by the models for the input sequences. This means that each input sequence is related to a fault scenario with the specified probability.

Both models properly predicted fault scenario 1. However, the model presented in ~\cite{no} failed to accurately predict fault scenario 2. It predicted irrelevant fault scenarios instead of predicting fault scenario 2 for each input sequence. However, the benefit of the method presented in ~\cite{no} over our approach is that this method extracts a graphical model of the process, which is beneficial for analyzing the process. The model extracted by the proposed method in this paper does not have such capability.

\begin{table*}[t]
\centering
\caption{Classification of the triggered alarms related to the first 15 occurrences of fault scenario 1 using the method presented in \cite{no}} 
\label{nm}
\begin{tabular}{@{}ccccc@{}}
\toprule
Fault 1
&
Fault 2
&
Fault 6
&
Fault 7
 &
Fault 11
\\
\midrule
1
&
0
&
0
&
0
&
0
\\
\hline
1
&
0
&
0
&
0
&
0
\\
\hline
1
&
0
&
0
&
0
&
0
\\
\hline
1
&
0
&
0
&
0
&
0
\\
\hline
1
&
0
&
0
&
0
&
0
\\
\hline
1
&
0
&
0
&
0
&
0
\\
\hline
1
&
0
&
0
&
0
&
0
\\
\hline
1
&
0
&
0
&
0
&
0
\\
\hline
1
&
0
&
0
&
0
&
0
\\
\hline
1
&
0
&
0
&
0
&
0
\\
\hline
1
&
0
&
0
&
0
&
0
\\
\hline
1
&
0
&
0
&
0
&
0
\\
\hline
1
&
0
&
0
&
0
&
0
\\
\hline
1
&
0
&
0
&
0
&
0
\\
\hline
1
&
0
&
0
&
0
&
0
\\
\bottomrule
\end{tabular}
\end{table*}

\begin{table*}[t]
\centering
\caption{Classification of the triggered alarms related to the first 15 occurrences of fault scenario 2 using the method presented in \cite{no}} 
\label{op}
\begin{tabular}{@{}ccccc@{}}
\toprule
Fault 1
&
Fault 2
&
Fault 6
&
Fault 7
 &
Fault 11
\\
\midrule
1
&
1
&
0
&
0
&
1
\\
\hline
0
&
1
&
0
&
0
&
1
\\
\hline
1
&
1
&
0
&
0
&
1
\\
\hline
0
&
1
&
0
&
0
&
1
\\
\hline
0
&
1
&
0
&
0
&
1
\\
\hline
0
&
1
&
0
&
0
&
1
\\
\hline
0
&
1
&
0
&
0
&
1
\\
\hline
0
&
0
&
0
&
0
&
1
\\
\hline
1
&
1
&
0
&
0
&
1
\\
\hline
1
&
1
&
0
&
0
&
1
\\
\hline
0
&
1
&
0
&
0
&
1
\\
\hline
0
&
0
&
0
&
0
&
1
\\
\hline
0
&
1
&
0
&
0
&
0
\\
\hline
1
&
0
&
0
&
0
&
1
\\
\hline
0
&
1
&
0
&
0
&
1
\\
\bottomrule
\end{tabular}
\end{table*}

\begin{table*}[t]
\centering
\caption{Classification of the triggered alarms related to the first 15 occurrences of fault scenario 1 using the method presented in this paper} 
\label{cas}
\begin{tabular}{@{}cccccccccc@{}}
\toprule
Fault 1
&
Fault 2
&
Fault 6
&
Fault 7
&
Fault 8
&
Fault 10
&
Fault 11
&
Fault 12
&
Fault 13
&
Fault 17
\\
\midrule
1
&
0
&
0
&
0
&
0
&
0
&
0
&
0
&
0
&
0
\\
\hline
1
&
0
&
0
&
0
&
0
&
0
&
0
&
0
&
0
&
0
\\
\hline
1
&
0
&
0
&
0
&
0
&
0
&
0
&
0
&
0
&
0
\\
\hline
1
&
0
&
0
&
0
&
0
&
0
&
0
&
0
&
0
&
0
\\
\hline
1
&
0
&
0
&
0
&
0
&
0
&
0
&
0
&
0
&
0
\\
\hline
1
&
0
&
0
&
0
&
0
&
0
&
0
&
0
&
0
&
0
\\
\hline
1
&
0
&
0
&
0
&
0
&
0
&
0
&
0
&
0
&
0
\\
\hline
1
&
0
&
0
&
0
&
0
&
0
&
0
&
0
&
0
&
0
\\
\hline
1
&
0
&
0
&
0
&
0
&
0
&
0
&
0
&
0
&
0
\\
\hline
1
&
0
&
0
&
0
&
0
&
0
&
0
&
0
&
0
&
0
\\
\hline
1
&
0
&
0
&
0
&
0
&
0
&
0
&
0
&
0
&
0
\\
\hline
1
&
0
&
0
&
0
&
0
&
0
&
0
&
0
&
0
&
0
\\
\hline
1
&
0
&
0
&
0
&
0
&
0
&
0
&
0
&
0
&
0
\\
\hline
1
&
0
&
0
&
0
&
0
&
0
&
0
&
0
&
0
&
0
\\
\hline
1
&
0
&
0
&
0
&
0
&
0
&
0
&
0
&
0
&
0
\\
\bottomrule
\end{tabular}
\end{table*}

\begin{table*}[h]
\centering
\caption{Classification of the triggered alarms related to the first 15 occurrences of fault scenario 2 using the method presented in this paper} 
\label{cat}
\begin{tabular}{@{}cccccccccc@{}}
\toprule
Fault 1
&
Fault 2
&
Fault 6
&
Fault 7
&
Fault 8
&
Fault 10
&
Fault 11
&
Fault 12
&
Fault 13
&
Fault 17
\\
\midrule
0
&
0.34
&
0
&
0
&
0
&
0
&
0.66
&
0
&
0
&
0
\\
\hline
0
&
0
&
0
&
0
&
0
&
0
&
0
&
0
&
0
&
0
\\
\hline
0
&
1
&
0
&
0
&
0
&
0
&
0
&
0
&
0
&
0
\\
\hline
0
&
1
&
0
&
0
&
0
&
0
&
0
&
0
&
0
&
0
\\
\hline
0
&
0.40
&
0
&
0
&
0.60
&
0
&
0
&
0
&
0
&
0
\\
\hline
0
&
1
&
0
&
0
&
0
&
0
&
0
&
0
&
0
&
0
\\
\hline
0
&
1
&
0
&
0
&
0
&
0
&
0
&
0
&
0
&
0
\\
\hline
0
&
1
&
0
&
0
&
0
&
0
&
0
&
0
&
0
&
0
\\
\hline
0
&
1
&
0
&
0
&
0
&
0
&
0
&
0
&
0
&
0
\\
\hline
0
&
1
&
0
&
0
&
0
&
0
&
0
&
0
&
0
&
0
\\
\hline
0
&
1
&
0
&
0
&
0
&
0
&
0
&
0
&
0
&
0
\\
\hline
0
&
0
&
0
&
0
&
0
&
0
&
0
&
0
&
1
&
0
\\
\hline
0
&
1
&
0
&
0
&
0
&
0
&
0
&
0
&
0
&
0
\\
\hline
0
&
0
&
0
&
0
&
0
&
0
&
0
&
0
&
1
&
0
\\
\hline
0
&
0&
0
&
0
&
0
&
0
&
0
&
0
&
1
&
0
\\
\bottomrule
\end{tabular}
\end{table*}

\section*{Conclusion}
Due to equipment failures, faults are inevitable in industrial processes. Alarm systems have been designed to raise the alarm when a fault occurs. So, alarms are guidelines for the faults. It is feasible to assist the operator in determining the root cause of alarms by modeling the relations between alarms using historical alarm data. 
 In this paper, a neural network model for online fault detection is proposed. This model was evaluated using the Tennessee Eastman process alarm data and used for online fault detection. 
 
The purpose of this paper is to present a neural network to classify alarm sequences considering three criteria. First, the amount of training data available in industrial processes is restricted. Second, time is a significant aspect in determining the underlying cause of an alarm in industrial processes. The model's performance is measured in terms of accuracy and precision. The higher these metrics, the better the model's performance for online implementation of the algorithm.
Hence, the proposed model should fulfill these three criteria. Furthermore, expert knowledge is not required for training the presented model. However, to use this method, alarm data must be labeled. The data used in this paper is simulation data, but industrial data is unlabeled and should primarily be labeled, which is not an easy task. In addition, the proposed method does not provide a graphical model of the relations between alarms. Thus, analyzing the relations between alarms is not possible.



\normalsize
\bibliography{refpaper}


\end{document}